\documentclass[a4paper,11pt]{article}
\pdfoutput=1

\usepackage{jcappub}
\usepackage[T1]{fontenc}
\usepackage{aas_macros}
\usepackage{bbm}
\usepackage[normalem]{ulem}
\usepackage{physics}
\usepackage{subcaption}

\newcommand{\lcdm}{\ensuremath{\Lambda\rm{CDM}}}
\newcommand{\omegam}{\ensuremath{\Omega_{\rm{m}}}}
\newcommand{\dlogz}{\ensuremath{\Delta\log{Z}}}
\newcommand{\dchisq}{\ensuremath{\Delta\chi^2}}
\newcommand{\one}{\ensuremath{\mathbbm{1}}} 
\newcommand{\diff}{\ensuremath{\mathrm{d}}} 

\definecolor{beaver}{rgb}{0.62,0.51,0.44}

\usepackage{scalerel}
\usepackage{tikz}
\usetikzlibrary{svg.path}

\definecolor{orcidlogocol}{HTML}{A6CE39}
\tikzset{
  orcidlogo/.pic={
    \fill[orcidlogocol] svg{M256,128c0,70.7-57.3,128-128,128C57.3,256,0,198.7,0,128C0,57.3,57.3,0,128,0C198.7,0,256,57.3,256,128z};
    \fill[white] svg{M86.3,186.2H70.9V79.1h15.4v48.4V186.2z}
                 svg{M108.9,79.1h41.6c39.6,0,57,28.3,57,53.6c0,27.5-21.5,53.6-56.8,53.6h-41.8V79.1z M124.3,172.4h24.5c34.9,0,42.9-26.5,42.9-39.7c0-21.5-13.7-39.7-43.7-39.7h-23.7V172.4z}
                 svg{M88.7,56.8c0,5.5-4.5,10.1-10.1,10.1c-5.6,0-10.1-4.6-10.1-10.1c0-5.6,4.5-10.1,10.1-10.1C84.2,46.7,88.7,51.3,88.7,56.8z};
  }
}

\newcommand\orcidicon[1]{\href{https://orcid.org/#1}{\mbox{\scalerel*{
\begin{tikzpicture}[yscale=-1,transform shape]
\pic{orcidlogo};
\end{tikzpicture}
}{|}}}}

\title{\boldmath Bayesian vs Frequentist: Comparing Bayesian model selection with a frequentist approach using the iterative smoothing method}

\author[a,b]{Hanwool Koo\orcidicon{0000-0003-0268-4488}}
\author[a,c]{Ryan E. Keeley\orcidicon{0000-0002-0862-8789}}
\author[a,b]{Arman Shafieloo\orcidicon{0000-0001-6815-0337}} 
\author[d]{and Benjamin L'Huillier\orcidicon{0000-0003-2934-6243}}

\affiliation[a]{Korea Astronomy and Space Science Institute (KASI),\\776 Daedeok-daero, Yuseong-gu, Daejeon 34055, Korea}
\affiliation[b]{KASI Campus, University of Science and Technology,\\217 Gajeong-ro, Yuseong-gu, Daejeon 34113, Korea}
\affiliation[c]{Department of Physics, University of California Merced, 5200 North Lake Road, Merced, CA 95343, USA}
\affiliation[d]{Department of Physics and Astronomy, Sejong University,\\209 Neungdong-ro, Gwangjin-gu, Seoul 05006, Korea}

\emailAdd{hkoo@kasi.re.kr}
\emailAdd{rkeeley@kasi.re.kr}
\emailAdd{shafieloo@kasi.re.kr}
\emailAdd{benjamin@sejong.ac.kr}

\abstract{We have developed a frequentist approach for model selection which determines the consistency between any cosmological model and the data using the distribution of likelihoods from the iterative smoothing method. Using this approach, we have shown how confidently we can conclude whether the data support any given model without comparison to a different one. In this current work, we compare our approach with the conventional Bayesian approach based on the estimation of the Bayesian evidence using nested sampling. We use simulated future Roman (formerly WFIRST)-like type Ia supernovae data in our analysis. We discuss the limits of the Bayesian approach for model selection and show how our proposed frequentist approach can perform better in the falsification of individual models. Namely, if the true model is among the candidates being tested in the Bayesian approach, that approach can select the correct model. If all of the options are false, then the Bayesian approach will select merely the least incorrect one. Our approach is designed for such a case and we can conclude that all of the models are false.}

\begin{document}
\maketitle
\flushbottom

\section{Introduction}\label{sec:intro}

The concordance model of cosmology, $\lcdm$ ($\Lambda$ for the cosmological constant and CDM for the cold dark matter), is facing conflicts and tensions. It has been the most successful model that explained various astronomical observations with remarkable simplicity and no significant change for decades. Though $\lcdm$ is consistent with low and high-redshift observations individually, it is in conflict with the combination of low and high-redshift data. This is the case with the $H_0$ tension, a discrepancy between the present expansion rate measured directly from the Cepheid calibration of Type Ia supernova (SN Ia) distances~\cite{riess-etal19} and that rate derived from the cosmic microwave background (CMB)~\cite{planck-etal20}. 

Using a model independent analysis to reconstruct the expansion history of the universe can be a useful approach to shed light on the current tensions, but the large gap in the data between the low (such as SN Ia and baryon acoustic oscillation) and high redshift observations (CMB) prevents us from having a complete reconstruction of the expansion history all way to the last scattering surface. Accepting our limitations, it is reasonable to look for alternative models that may perform better than the standard model at fitting combinations of cosmological observations, as well as falsifying such models in a robust manner. 

In our previous paper we introduced a frequentist approach~\cite{koo-etal21} to test the consistency between any individual cosmological model and the SN Ia observations which are one of the most reliable cosmological datasets. This method is based on the non-parametric iterative smoothing method, introduced and improved by \cite{shafieloo-etal06,shafieloo07,shafieloo-clarkson10,shafieloo-etal18}, which reconstructs the distance modulus in a model-independent way. Our previous paper provides a detailed description about our method, about how to generate and use the likelihood distributions based on the data covariance matrix. We also showed that our method can also be used to perform parameter estimation for each model. 

In this work we elaborate further on the important subject of model selection and we compare the performance of our proposed method with that of the conventional approach using Bayesian statistics, namely the Bayesian evidence ratio. To perform this analysis, we simulated a mock Nancy Grace Roman Space Telescope~\cite[Roman, formerly WFIRST,][]{green-etal12} SN Ia dataset. We will show that, using our approach, not only can we successfully distinguish and make a preference among different alternative models but we can also rule out all alternative candidates if none is the correct model. This is a great advantage over the conventional Bayesian evidence approach which, in the scenario where none of the candidate models is the true one, will still select one of the false models as best and thus not rule it out.

We concretely demonstrate these points by considering three different dark energy models. We imagine a case where the true model, the model which generated the data, is an unknown unknown; it is not in the list of models that are considered possibilities. Rather only two false models are considered. We intend to demonstrate that the standard Bayesian model selection approach will be able to tell which of the two models is the  preferred or the less bad fit, but will not be able to identify that neither models are true. In turn, we intend to show that our methodology can identify that both false models are indeed false.

We introduce the methodology of using the Bayesian evidence for model selection in Section~\ref{sec:dlogz}, and of using the iterative smoothing method and likelihood distributions in Section~\ref{sec:dchi2}. We present our results in Section~\ref{sec:res} and summary in Section~\ref{sec:sum}.

\section{Model selection using Bayesian evidence}\label{sec:dlogz}

In this section, we discuss the common method for comparing models by estimating the Bayesian evidence. The Bayesian evidence $Z$, also called marginal likelihood, is defined as the integral of the product of the likelihood $L(\theta)$ and the prior $\pi(\theta)$ taken over the entire parameter space of $\theta$:
\begin{equation}\label{eqn:evidence}
Z=\int L(\theta)\pi(\theta) \,\diff \theta.
\end{equation}
The Bayesian evidence is the average of the likelihood of the data over the parameter space. One notable feature of the Bayesian evidence is that it penalizes extended parameter spaces that do not fit the data well. That is, the parameter space that is within the prior but not within the peak of likelihood will decrease the evidence. In other words, the larger the fraction of the parameter space that is occupied by the peak of the likelihood the larger is the evidence that results. Thus when comparing two models' evidences, the evidence gives advantage to the model with less wasted parameter space i.e. the model that is more predictive.

The Bayes factor, which is the ratio of the evidence of the two different models, is one of the most reliable model selection manner in Bayesian statistics. If the Bayes factor is larger than unity, or equivalently, if the difference in the log of the evidences of Model 1 ($\rm M_2$) and Model 2 ($\rm M_1$)
\begin{equation}\label{eqn:dlogz}
\dlogz =\log{Z(\rm M_1)} - \log{Z(\rm M_2)}
\end{equation}
is positive, $\rm M_1$ is supported by the data over $\rm M_2$. We can measure the strength of the preference of $\rm M_2$ over $\rm M_1$ using the interpretation of $\dlogz$ shown in Table~\ref{tab:scale_dlogz}. This scale, the Kass-Raftery (KR) scale was suggested by \cite{kass-raftery95} and is a more conservative modification of the widely used Jeffreys' scale~\cite{jeffreys39}.\footnote{Previous studies including~\cite{nesseris-garciabellido13} have discussed reliability of these scales.} Based on this interpretation, we forecast how frequently the analysis using Bayesian evidence will strongly support $\rm M_1$ when we simulate future SN Ia datasets and calculate the distribution of $\dlogz$.

\begin{table}[htbp]
\centering
\begin{tabular}{|c|c|}
\hline
$\dlogz$ & Evidence against $\rm M_1$\\
\hline
0 to 1 & Negligible \\
1 to 3 & Positive \\
3 to 5 & Strong \\
 $>$ 5 & Very strong \\
\hline
\end{tabular}
\caption{\label{tab:scale_dlogz} The Kass-Raftery scale: a conservative interpretation of $\dlogz$ for model selection~\cite{kass-raftery95}. }
\end{table}

We use \texttt{mc3}~\cite{cubillos-etal17} and \texttt{dynesty}~\cite{speagle20}, which implements a dynamic nested sampling~\cite{higson-etal19} algorithm, to calculate the Bayesian evidence. Nested sampling~\cite{skilling04,skilling06} estimates the Bayesian evidence by transforming the integral over the N-dimensional parameter space into a one-dimensional integral over the prior volume that is contained within an iso-likelihood surface. Numerically, the evidence is the weighted sum of likelihood values; the weights are determined by the nested sampling algorithm. The algorithm repeatedly removes the point with the lowest likelihood among multiple `live' points that represent different iso-likelihood surfaces, and replace it with a new live point with higher likelihood that is chosen from the Monte Carlo samples. For each iteration, live points are used for estimating prior volume of the previous live point, then we numerically calculate the evidence by summing the product of the likelihoods and prior volumes of all previous live points. The algorithm stops when the calculated evidence converges to within a specified error tolerance. It is called dynamic nested sampling when the number of effective live points is variable at any given iteration. This variable number of live points allows for modulating the speed at which the algorithm converges to the integral.

\section{Model selection using the iterative smoothing method and likelihood distributions}\label{sec:dchi2}

In this section, we discuss our frequentist approach for model selection, which uses the likelihood distribution from the iterative smoothing method. 
Our iterative smoothing technique~\cite{shafieloo-etal06,shafieloo07,shafieloo-clarkson10,shafieloo-etal18} seeks to reconstruct a smooth function of the distance modulus $\mu(z)$ from the data $\mu_i$ observed at redshifts $z_i$. This method starts from an arbitrary initial guess, $\hat{\mu}_0(z)$ and perturbs this guess such that the residuals look smoother and more Gaussian. This procedure works iteratively such that ($n+1$)th iteration, $\hat{\mu}_{n+1}(z)$, is calculated by
\begin{equation}\label{eqn:smooth}
\hat{\mu}_{n+1}(z) = \hat{\mu}_n(z) + \frac{\boldsymbol{\delta\mu_n}^T \cdot \mathbf{C^{-1}} \cdot \boldsymbol{W}(z)}{\one^T \cdot \mathbf{C^{-1}} \cdot \boldsymbol{W}(z)},
\end{equation}
where $\one^T=(1,\cdots,1)$, the weight $\boldsymbol{W}$ and residual $\boldsymbol{\delta\mu_n}$ are
\begin{equation}\label{eqn:weight}
W_i(z)=\exp\left[-\frac{\ln^2\left(\frac{1+z}{1+z_i}\right)}{2\Delta^2}\right]
\end{equation}
\begin{equation}\label{eqn:residual}
\boldsymbol{\delta\mu_n}|_i = \mu_i -  \hat{\mu}_n(z_i)
\end{equation}
and $\mathbf{C^{-1}}$ indicates the inverse of covariance matrix of the data. The smoothing width is set to $\Delta = 0.3$ following previous analyses in \cite{shafieloo-etal06, lhuillier-shafieloo17, lhuillier-etal18, koo-etal20}. We define the $\chi^2$ value of the reconstruction $\hat{\mu}_n(z)$ as
\begin{equation}\label{eqn:chi2}
\chi_n^2 = \boldsymbol{\delta\mu_n}^T \cdot \mathbf{C^{-1}} \cdot \boldsymbol{\delta\mu_n}.
\end{equation}
By design, the iterative smoothing method reconstructs a function which at any iteration fits the data better than at the previous iteration. Furthermore, after a large number of iterations the reconstructions converge to a unique solution independent of the choice of the initial guess. In other words, when starting from different initial guesses, the likelihood of the first iterations may be very different, but after a large number of iterations, the final reconstructions converge to the same solution with a unique likelihood. We use the 1000th iteration of the iterative smoothing method, which is large enough to achieve this convergence (that generally occurs after a few hundreds of iterations)~\cite{shafieloo07,lhuillier-etal18,shafieloo-etal18,koo-etal20}. 

In the previous work~\cite{koo-etal21}, we have applied the smoothing procedure to derive the distribution of the difference in $\chi^2$ between the smoothed function and that of the best-fit of the model being tested ($\dchisq = \chi^2_{\rm smooth} - \chi^2_{\rm best-fit}$). We call this distribution the likelihood distribution, which follows a frequentist statistical approach and shows how frequently the procedure generates better fits by certain amount of improvement. Then we posed the question: ``how much better does this improvement have to be in order to be significant?'' and answered it by finding the $\dchisq$ value such that only $5\%$ or $1\%$ of the time would the smoothing method randomly achieve a better improvement than this value. In other words, we have determined the $\dchisq$ values that correspond to the $95\%$ and $99\%$ confidence limits (CLs), $\dchisq_{95\%}$ and $\dchisq_{99\%}$. We also demonstrated that likelihood distribution and the corresponding $95\%$ and $99\%$ CLs are independent of the assumed cosmological model. If the $\dchisq$ of real data is larger than the $95\%$ (99\%) CLs, then we can conclude that the model is inconsistent with the data at 2$\sigma$ (3$\sigma$) significance, even though the true model is unknown.

In this work, we perform model selection by deriving the likelihood distribution using 1000 mock realizations from Roman covariance matrix and previously calculated values of $\dchisq_{95\%}$ and $\dchisq_{99\%}$. We forecast how frequently each model is ruled out by the simulated future SN Ia datasets and compare the result with that of the conventional Bayesian analysis.

\section{Data simulations \& implementation of methods}\label{sec:res}

SN Ia distance measurements have become one of the most important datasets of modern cosmology. Since they are standardizable candles, we can use them to directly measure the accelerating expansion of the Universe at late times~\cite{riess-etal98,perlmutter-etal99}. Almost all previous SN Ia compilations including SuperNova Legacy Survey~\cite[SNLS,][]{sullivan-snlsc05}, Gold~\cite{riess-etal07}, Union~\cite{kowalski-etal08}, Constitution~\cite{hicken-etal09}, Union2~\cite{amanullah-etal10}, Union2.1~\cite{suzuki-etal12}, Joint Light-curve Analysis~\cite[JLA,][]{betoule-etal14} and Pantheon~\cite{scolnic-etal18} have been shown to be consistent with the flat $\lcdm$ model. However, these consistency tests require some assumptions for parametrization or functional form. There have also been some model-independent analyses that search for systematics and test the internal consistencies in a non-parametric manner~\cite{lhuillier-etal19,koo-etal20,keeley-etal21}.

To generate simulated future SN Ia data, we first need to construct a covariance matrix. We do this by constructing its diagonal term using the forecasted error information of the Roman telescope
 that the total error in luminosity distance is $\sim 0.5$-$1\%$ between $z=0.1$-$1.7$~\cite{spergel-etal15}. With a covariance matrix in hand, we can then generate Gaussian random mock datasets based on an assumed fiducial model (here the Transitional Dark Energy (TDE) model~\cite{keeley-etal19}). We use the TDE model with the Hubble constant $H_0=70$ km s$^{-1}$ Mpc$^{-1}$ and matter density $\omegam=0.3$ as the fiducial model and generate 1000 mock realizations. Then, we choose two other models, the $\lcdm$ and Phenomenologically Emergent Dark Energy (PEDE)~\cite{li-shafieloo19,li-shafieloo20} models, as candidates in the model selection procedures under discussion (the Bayesian evidence ratio approach and our proposed approach based on the iterative smoothing method).

\begin{figure*}[tbp]
\centering
\begin{subfigure}[b]{.495\linewidth}
\includegraphics[width=\textwidth]{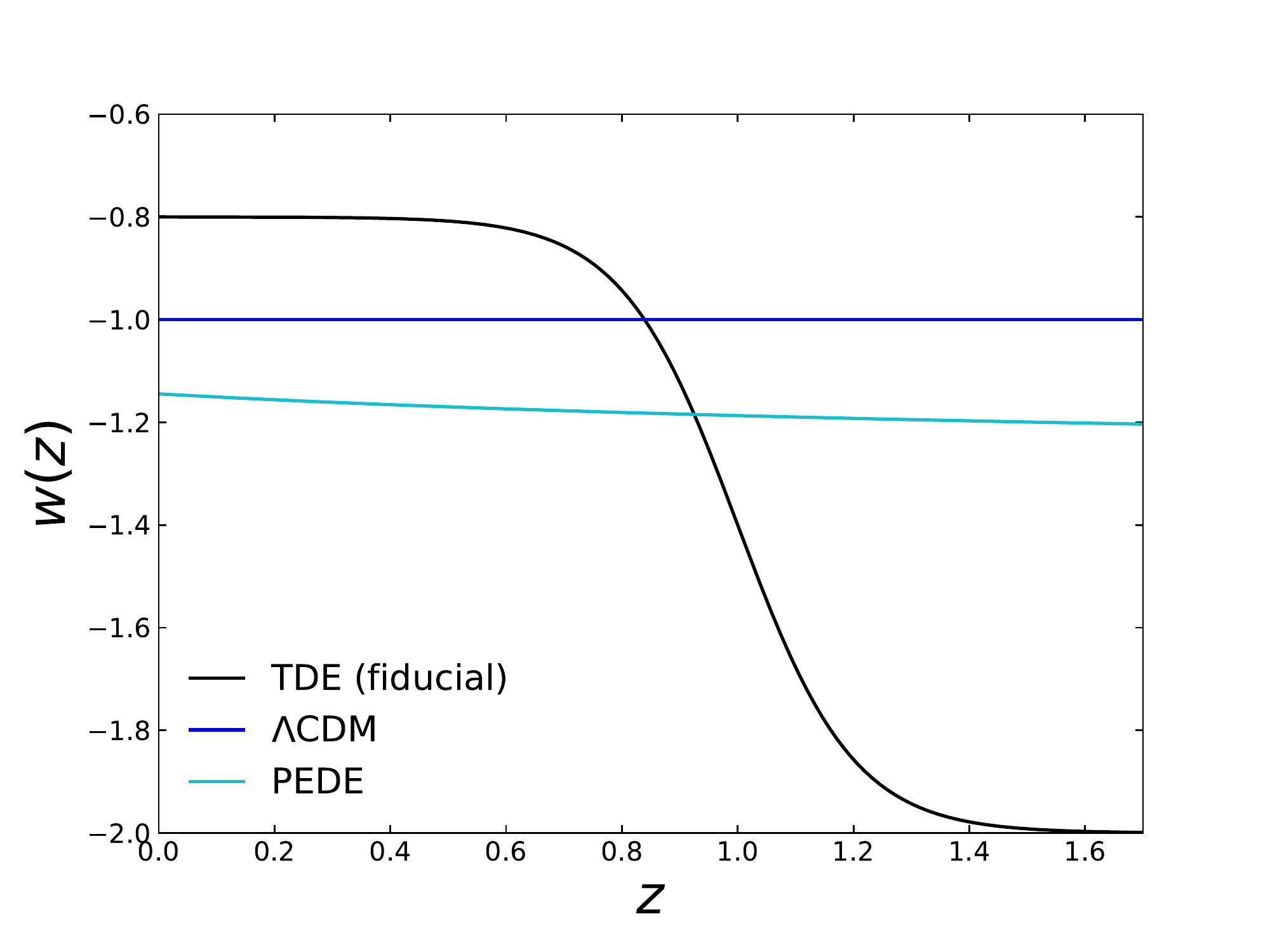}
\subcaption{\label{fig:w}Equation-of-state parameter}
\end{subfigure}
\begin{subfigure}[b]{.495\linewidth}
\includegraphics[width=\textwidth]{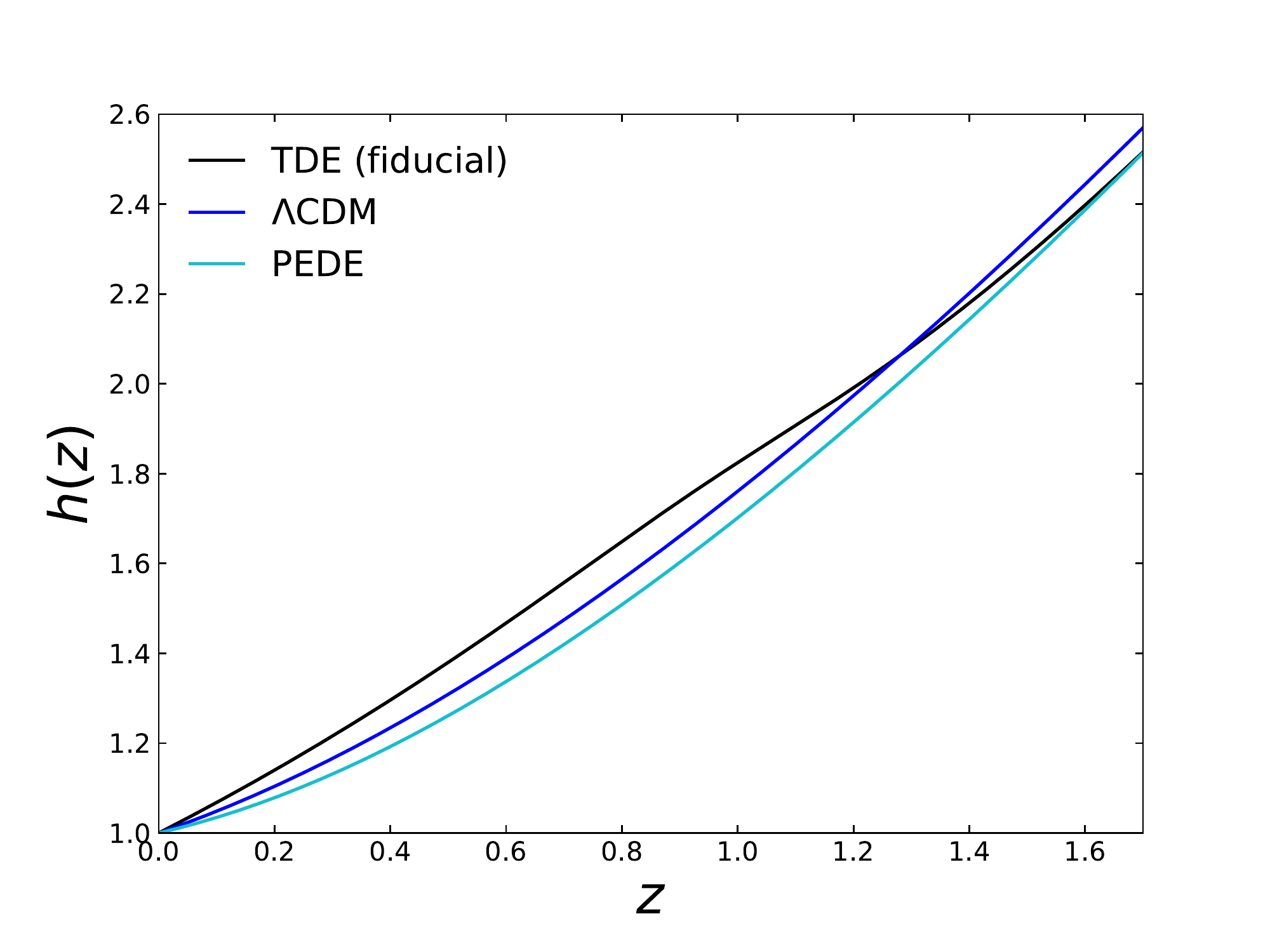}
\subcaption{\label{fig:h}Expansion history}
\end{subfigure}
\caption{\label{fig:param} The equation-of-state parameter $w(z)$ (left) and expansion history $h(z)$ (right) for the TDE (black; fiducial) model and two comparison models, $\lcdm$ (blue) and PEDE (cyan), with the same values of $\omegam=0.3$.}
\end{figure*}

The difference between these models ($\lcdm$, PEDE, TDE) is the difference between their equations-of-state parameters. The equation-of-state parameter is $w(z)=-1$ for the $\lcdm$ model and
\begin{equation}\label{eqn:w_pede}
    w(z) = -\tfrac{1}{3\ln 10} \left({1+\tanh\left[\log_{10}\,(1+z)\right]}\right)-1
\end{equation}
for PEDE model. For the TDE model, it is parametrized as
\begin{equation}\label{eqn:w_tde}
w(z) = w_0 + (w_1 - w_0)\left(1 + \tanh(\tfrac{z - z_t}{\Delta_z})\right)/2
\end{equation}
where we choose $(w_0,\ w_1,\ z_t,\ \Delta_z) = (-0.8,\ -2.0,\ 1.0,\ 0.2)$. These values were chosen to make an example model that is very different than PEDE or $\lcdm$. We can see these differences in the equation-of-state parameter in Figure~\ref{fig:w}. In contrast with PEDE, which is purely phantom, TDE starts out in the quintessence regime at lower redshift (where most of the data is) but then transitions to the phantom regime at higher redshift. We fix the parameters that determine the equation of state in the TDE model to give it the same number of degrees of freedom as the $\lcdm$ model.

Next, we calculate the expansion history of the Universe, $h(z) \equiv H(z)/H_0$ where $H(z)$ is the Hubble parameter, for all these models via
\begin{equation}\label{eqn:h_z}
h^2(z) = \omegam(1+z)^3 + (1-\omegam)\exp(3\int_{0}^{z}\frac{1+w(z')}{1+z'}dz').
\end{equation}
Figure~\ref{fig:h} shows the results of this calculation of the expansion history of the Universe for the $\lcdm$, PEDE and fiducial TDE model, with the same values of $\omegam=0.3$. It demonstrates how, despite the equation-of-state parameter being very different between the considered models, the expansion histories of the different models are relatively close to each other since they involve integrals of the equation of state. For each of these models, we allow $\omegam$ and $H_0$ to vary within $[0.0,\ 1.0]$ and $[60,\ 80]$ km s$^{-1}$ Mpc$^{-1}$.

\subsection{Results from Bayesian analysis}\label{ssec:bayes}

\begin{figure}[tbp]
\centering
\includegraphics[width=.495\textwidth]{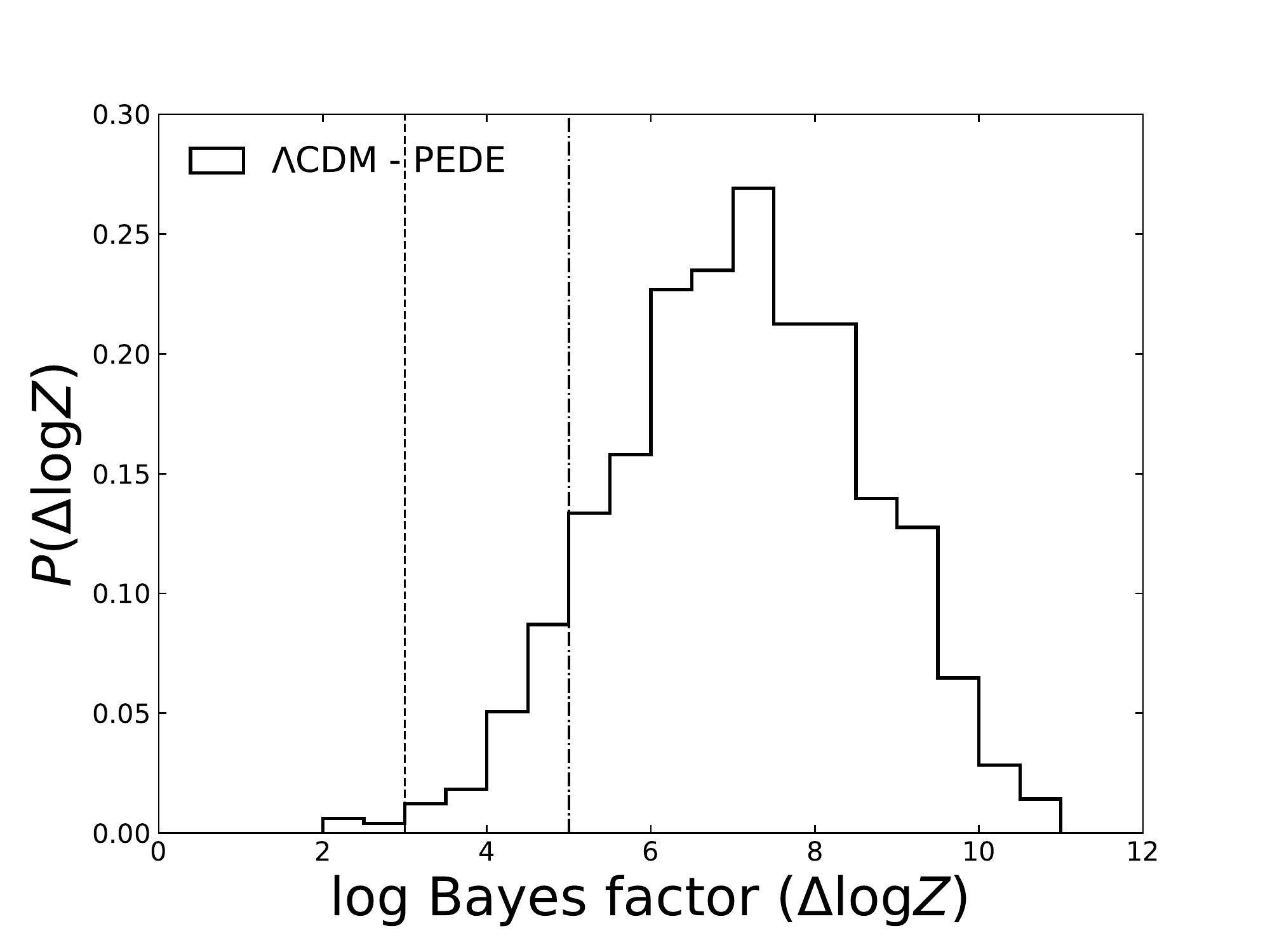}
\hfill
\caption{\label{fig:dlogz_2models} Distributions of $\dlogz$ for each of the considered pairs of models between the $\lcdm$ and PEDE model. Vertical lines correspond to $\dlogz=3$ and $5$. The distribution is over the different realizations of mock Roman datasets based on the fiducial TDE model. One can see that the Bayesian evidence approach strongly favors $\lcdm$ over PEDE, while in fact both of these models are false models.} 
\end{figure}

\begin{table}[tbp]
\centering
\begin{tabular}{|c|cc|}
\hline
$\dlogz >3$ & PEDE consistent& PEDE ruled-out\\
\hline
$\lcdm$ consistent & 6 & 994\\
$\lcdm$ ruled-out & 0 & 0\\
\hline
\end{tabular}
\begin{tabular}{|c|cc|}
\hline
$\dlogz >5$ & PEDE consistent & PEDE ruled-out\\
\hline
$\lcdm$ consistent & 89 & 911\\
$\lcdm$ ruled-out & 0 & 0\\
\hline
\end{tabular}
\caption{\label{tab:dlogz_2models} The number of realizations among 1000 realizations that strongly and very strongly supports the $\lcdm$ model against PEDE model from the analysis using the Bayesian evidence. The data for each of these cases are generated from the TDE model and thus none of the two models in the table are correct models.}
\end{table}

We present the results of our Bayesian analysis in Figure~\ref{fig:dlogz_2models}. There is shown the distribution of $\dlogz$ between the $\lcdm$ and PEDE models. The distribution is over the different realizations of mock Roman-like datasets based on the fiducial TDE model. It is a testament to the precision of the mock Roman-like datasets that the Bayesian evidence gives ``Positive evidence'' that supports $\lcdm$ over PEDE in all of the cases, ``Strong evidence'' in $99.4\%$ of the cases and ``Very strong evidence'' in $91.1\%$ of the cases, as given from Table~\ref{tab:dlogz_2models} in the KR scale. In other words, the Bayesian evidence strongly supports $\lcdm$ over PEDE in almost all of the mock datasets, yet neither of these models is the correct model. In no cases are both incorrect models found to be inconsistent with the data. Both $\lcdm$ and PEDE have the same degrees of freedom, and as already shown at low redshift in Figure~\ref{fig:param}, the expansion history of the fiducial TDE model is closer to that of $\lcdm$ than that of PEDE. This explains why the conventional Bayesian analysis supports the $\lcdm$ model over PEDE even though neither are correct.

\subsection{Results from frequentist analysis}\label{ssec:freq}

\begin{figure}[tbp]
\centering 
\includegraphics[width=.495\textwidth]{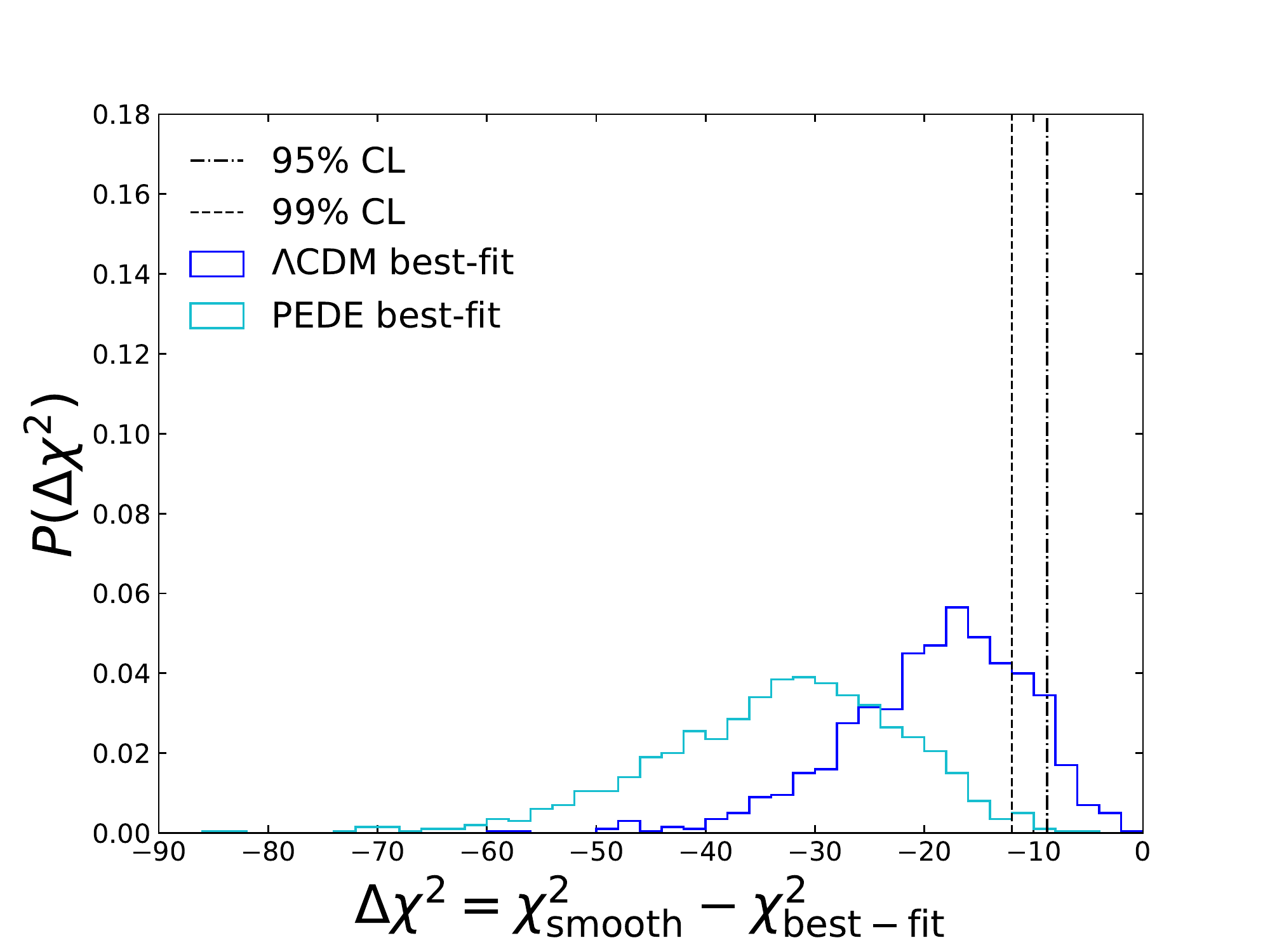}
\hfill
\caption{\label{fig:dchi2_2models} Likelihood distributions of $\dchisq = \chi^2_{\rm smooth} - \chi^2_{\rm best-fit}$ for each of the considered models, $\lcdm$ (blue) and PEDE (cyan), from the analysis using the iterative smoothing method. Vertical lines show the $95\%$ (dashed) and $99\%$ (dash-dotted) CLs of the likelihood distribution. These limits represent the expected amount of improvement (over-fitting) the smoothing procedure will perform on the true model. The distribution is for 1000 mock realizations of Roman dataset based on the fiducial TDE model. We can see that, using this approach, in a large number of cases, both $\lcdm$ and PEDE would be ruled out.}
\end{figure}

\begin{table}[tbp]
\centering
\begin{tabular}{|c|cc|}
\hline
95\% CL & PEDE consistent& PEDE ruled-out\\
\hline
$\lcdm$ consistent & 2 & 82\\
$\lcdm$ ruled-out & 0 & 916\\
\hline
\end{tabular}
\begin{tabular}{|c|cc|}
\hline
99\% CL & PEDE consistent & PEDE ruled-out\\
\hline
$\lcdm$ consistent & 14 & 193\\
$\lcdm$ ruled-out & 0 & 793\\
\hline
\end{tabular}
\caption{\label{tab:dchi2} The number of realizations among 1000 realizations within or outside the $95\%$, $99\%$ CLs for each model best-fits from the analysis using the iterative smoothing method. The data for each of these cases are generated from the TDE model and thus none of the two models in the table are correct models.}
\end{table}

To compare with the results from the Bayesian analysis, we calculate the likelihood distribution ($\dchisq$) for the $\lcdm$ and PEDE models. Figure~\ref{fig:dchi2_2models} displays these distributions. Also shown is the $95\%$ ($\dchisq_{95\%}$) and $99\%$ ($\dchisq_{95\%}$) CLs of the likelihood distribution. Table~\ref{tab:dchi2} shows the number of mock realizations among 1000 realizations that exceed or do not exceed the two different CLs for the $\lcdm$ and PEDE models. The number of realizations that exceeds the $95\%$, $99\%$ CLs is 916, 793 for $\lcdm$ and 998, 986 for PEDE. It is clear that using this approach, in $91.6\%$ of the cases both assumed models are ruled out at $95\%$ confidence and in $79.3\%$ of the cases they are both ruled out at $99\%$ confidence.

In both results, the data find the $\lcdm$ model to be favored over the PEDE model. The Bayes factors always support the $\lcdm$ model over PEDE with $\dlogz \geq 2$, which is at least ``Positive evidence'' according to the KR scale, and in our likelihood distribution analysis, Table~\ref{tab:dchi2} shows that in none of the cases where $\lcdm$ is found to be inconsistent with the data the PEDE model found to be consistent. However, the most obvious and important difference between the results is that our frequentist analysis can rule out both false models at 99\% confidence around 80\% of the time whereas the Bayesian analysis can never do so. We should emphasize here that the conventional Bayesian approach ranks the relative performances of different models with respect to the data but cannot absolutely rule out a model, while it is a very important advantage that our frequentist approach can rule out different candidates without any prior knowledge about the true model.

\section{Summary}\label{sec:sum}

In this paper, we expand upon the frequentist model selection method that previously introduced in~\cite{koo-etal21} and compare the statistical power of our test to the conventional approach based on the Bayesian evidence ratio. Our method works by calculating the likelihood distribution $\dchisq$ and using the values that enclose $95\%$ and $99\%$ of the volume of this distribution as criteria to test whether real data support a given model, independent of how well other models perform. The comparison of these two approaches is done using mock forecasted future Roman data based on TDE model.

Using the conventional Bayesian approach, we show that we can rule out the PEDE model in favor of $\lcdm$ model in $91.1\%$ of the cases with very strong evidence and in $99.4\%$ of the cases with strong evidence. However, there is no mock dataset for which the $\lcdm$ model can be ruled out by the method. On the other hand, we show that, using our frequentist iterative smoothing approach, we can rule out both competing candidate models ($\lcdm$ and PEDE) in $79.3\%$ of the cases with more than $99\%$ confidence and in $91.6\%$ of the cases with more than $95\%$ confidence. The method also shows that $\lcdm$ is closer to the fiducial TDE model than PEDE is since in $19.3\%$ ($8.2\%$) of our mock datasets, PEDE is ruled out while $\lcdm$ is found to be consistent at the $99\%$ ($95\%$) CL. Further, there is no mock dataset for which the $\lcdm$ model can be ruled out while PEDE is consistent with the dataset. Taken together, these results serve as a forecast that future Roman data should have enough SN Ia at high redshift to be able to not only distinguish models confidently using a Bayes factor procedure but also rule out individual models using just the data irrespective of how well other models perform.

These results indicate that our model selection methodology that uses the likelihood distribution tests individual models and can \emph{absolutely} rule out both false models. However model selection using the Bayesian evidence can only assess the \emph{relative} performance of the two models, but is not able to rule out a least-bad but still false model. This is possible since reconstructions generated by the iterative smoothing method provides model-independent information without knowledge of the true model and without the need for comparing both models with one another. It is a clear advantage of our non-parametric and frequentist approach over Bayesian methods for model selection when the true model is unknown as is the case in cosmology.

The analysis using realizations of Roman mock datasets can be done in the same way for forecasting results from other future SN Ia compilations, such as the ones from the Ten-year Rubin Observatory Legacy Survey of Space and Time~\cite[LSST,][]{ivezic-etal19}. These surveys may provide more SN Ia which can help us detect any possible deviation from the concordance $\lcdm$ cosmological model. Similarly we can use the same approach and philosophy to construct a model independent approach for model falsification, model selection and parameter estimation using other cosmological data. This will be our future work.

\acknowledgments

We thank Eric Linder, Savvas Nesseris, Leandros Perivolaropoulos and Wuhyun Sohn for useful comments. This work was supported by the high performance computing clusters Seondeok at the Korea Astronomy and Space Science Institute. A.~S. and B.~L. would like to acknowledge the support of the Korea Institute for Advanced Study (KIAS) grant funded by the government of Korea. B.~L. would like to acknowledge the support of the National Research Foundation of Korea (NRF-2019R1I1A1A01063740).

\bibliographystyle{JHEP}
\bibliography{KKSL21}

\end{document}